\documentclass{iucr}              

     \journalcode{J}              
\usepackage{layouts}
\usepackage{amsmath}
\usepackage{xcolor}
\begin{document}                  



\title{Ultimate Sensitivity in X-ray Diffraction: Angular Moments vs. Shot Noise}


\cauthor[a,b]{Peter}{Modregger}{peter.modregger@uni-siegen.de}{}
\author[a,b]{Felix}{Wittwer}
\author[a,b]{Ahmar}{Khaliq}
\author[b,c]{Niklas}{Pyrlik}

\author[d]{James A. D.}{Ball}

\author[e]{Jan}{Garrevoet}
\author[e]{Gerald}{Falkenberg}

\author[f]{Alexander}{Liehr}

\author[b]{Michael}{Stuckelberger}

\aff[a]{Physics department, University of Siegen, \country{Germany}}
\aff[b]{Centre for X-ray and Nanoscience, CXNS, DESY, Hamburg \country{Germany}}
\aff[c]{Physics department, University of Hamburg, \country{Germany}}
\aff[d]{European Synchrotron Radiation Facility (ESRF), Grenoble, \country{France}}
\aff[e]{Deutsches Elektronen-Synchrotron DESY, Hamburg, \country{Germany}}
\aff[f]{Institute of Mechanical Engineering, University of Kassel, \country{Germany}}








\newcommand{\zeroth}{0$^{\mathrm{th}}${}}
\newcommand{\first}{1$^{\mathrm{st}}${}}
\newcommand{\second}{2$^{\mathrm{nd}}${}}
\newcommand{\fourth}{4$^{\mathrm{th}}${}}

\maketitle                        


\begin{abstract}
The sensitivity of x-ray diffraction experiments towards Bragg peak parameters constitutes a crucial performance attribute of experimental setups. Frequently, diffraction peaks are characterized by model-free angular moment analysis, which offers a greater versatility compared to traditional model-based peak fitting. Here, we have determined the ultimate sensitivity of angular moments for diffraction data that is limited by photon shot noise. We report experimentally achieved sensitivities of the first moment below $1/1000$th of a detector pixel and below $1\mu$rad. We have demonstrated the validity of our theoretical predictions by an excellent agreement with experimental results from three different setups. The provided formulas for the uncertainties of angular moments allow for the rapid determination of experimentally achieved sensitivities from single diffraction frames.
\vspace{10mm}
\end{abstract}


\section{Introduction}

Indisputably, Bragg peak analysis plays an essential role in  x-ray diffraction experiments. The different parameters of diffraction peaks give insight to a large variety of sample properties, including material composition and identification of phases (via peak intensities), lattice parameters or internal strain and stress (via angular peak positions) or local crystal quality (via angular peak width)~\cite{SedighRahimabadi2020,Willmott2019}. In standard data analysis, these parameters are extracted from one-dimensional diffraction data by fitting to an appropriate model function. Some examples are Gaussian, Lorentzian or (Pseudo)-Voigt functions~\cite{Harrington2021}.

However, if the used model function deviates noticeably from the shape of the diffraction signal (e.g., double peaks or asymmetry), the fitted results can be ambiguous or misleading. Model-free peak analysis based on angular moments (also called centroid or center-of-mass method) do not face this challenge and are, thus, more and more commonly used~\cite{Liu2022,Ferrer2023,Sj2025}. For at least one possible application, i.e. the possibility for strain tensor tomography with x-ray diffraction, peak data analysis based on angular moments is mandatory~\cite{Lionheart2015}. Moment analysis is also used for other methods, where examples include  scanning small angle x-ray scattering~\cite{Bunk2009} or sub-pixel x-ray scattering~\cite{Modregger2017}.

An additional development in recent years is the rise of single photon counting detectors at synchrotron radiation beamlines dedicated to x-ray diffraction~\cite{Leu2016,Seo2019,Wright2020,Lu2021,Chakrabarti2022,Blankenburg2023,Stone2023}. This type of detector offer a unique combination of high sensitivity, a high dynamic range, close to absent readout and dark current noise; some detector models also offer limited photon energy resolution~\cite{Trueb2015}. By counting single photons, these detectors cannot avoid photon shot noise, which obeys the well-known Poisson statics~\cite{Willmott2019}. 

In the following, we will demonstrate that the known statistics of single photon detectors allows for a theoretical prediction of the sensitivities of the first three angular moments of one-dimensional diffraction peaks. We will validate the theory by comparison with experimental results from three different setups.

\section{Theory}

In this section we will derive the uncertainties $u$ due to photon shot noise of the first three moments $M_0$, $M_1$ and $M_2$ of a measured intensity distribution $f$ using error propagation. 

Suppose that a one-dimensional intensity distribution $f$ is measured as photon counts at equidistant sampling points $x_j$ with a separation of $\Delta x$. The \zeroth moment is given as~\cite{Frederik}:
\begin{equation}\label{eq:M0}
    M_0 = \sum_j f_j\, \Delta x
\end{equation}
with $j$, the number of the sampling position; the normalized \first moment is
\begin{equation}\label{eq:M1}
    M_1 = \sum_j x_j f_j\, \Delta x \, / \,  M_0;
\end{equation}
and, finally, the normalized and centralized higher moments of order $n$ are
\begin{equation}\label{eq:Mn}
    M_n = \sum_j (x_j-M_1)^n f_j\, \Delta x \, / \, M_0.
\end{equation}
These moments are well defined and distinct for any experimentally obtained function $f$. For sufficiently large sampling intervals, small sampling distances and in the absence of a background signal, the moments represent information about the underlying diffraction curve. Specifically, the \zeroth moment corresponds to the integrated photon count $N_{\mathrm{photons}}$, i.e. $M_0 / \Delta x = N_{\mathrm{photons}}$. The \first moment constitutes the center-of-mass (or centroid) position, which is identical to the peak position only for symmetric functions with a single peak. Otherwise, in some situations it can be useful to regard the \first moment as the weighted sum of many contributing symmetric single peaks~\cite{Hauk1997}. The square root of the \second moment corresponds to the width of a peak shape function. For the frequently occurring case of a Gaussian function the \second moment is identical with the square its standard deviation, i.e. $\sigma^2=M_2$.

The uncertainty of the photon count measurements $f_j$ due to photon shot noise are given by~\cite{Willmott2019}
\begin{equation}\label{eq:photon_shot_noise}
    u(f_j) = \sqrt{f_j}.
\end{equation}
Each individual intensity measurement $f_j$ is affected by shot noise and the overall impact on the moments can be calculated using error propagation~\cite{Barlow1989}
\begin{equation}\label{eq:error_propagation}
   u(M_n)^2 = \sum_j \left( \frac{\partial M_n}{\partial f_j} u(f_j)\right)^2.
\end{equation}
Applying this to the definition of moments (i.e., eqs.~(\ref{eq:M0})-(\ref{eq:Mn})) while considering the appropriate uncertainties, i.e. eq.~(\ref{eq:photon_shot_noise}) it is straightforward to show that the uncertainties of the first three moments subjected to photon shot noise are given as
\begin{eqnarray}
    u(M_0) &=& \sqrt{\Delta x\, M_0} \label{eq:uM0}\\
    u(M_1) &=& \sqrt{\Delta x\, M_2 \, / \, M_0} \label{eq:uM1} \\
    u(M_2) &=& \sqrt{\Delta x\, (M_4-{M_2}^2) \,/\, M_0} \label{eq:uM2}.
\end{eqnarray}
Details of the derivation are provided in the appendix. The results following intuition: The \zeroth moment constitutes the integrated photon counts and its absolute uncertainty increases while its relative uncertainty (i.e., $u(M_0) / M_0$) decreases with the number of collected photons; the uncertainty of the \first moment decreases with the number of photons and increases with the width of the peak; The uncertainty of the \second moment again decreases with the photon number while it increases with the difference between the \fourth and the square of the \second moment. Please note that this difference is always positive as $u(M_2)^2$ can be rewritten as $\Delta x\, (\kappa-1) \,/\, (M_0 {M_2}^2)$ with the kurtosis $\kappa$, which obeys $\kappa \ge 1$~\cite{Frederik}.

Evidently, experimentally encountered uncertainties maybe larger than these results due to x-ray beam intensity fluctuations (e.g intensity decay or injection), setup instabilities (e.g., drift of components) or non-linear response from the photon counting detectors. Some of these influences have been studied in the following experimental section. However, photon shot noise constitutes a fundamental limit, which render the retrieved uncertainties the ultimate achievable sensitivities.

\section{Results}

The basic idea to determine experimental uncertainties follows~\cite{Modregger2011a}: Measurements are repeated a number of times and the standard deviation of resulting values are taken as uncertainties. We want to highlight that the experimental details are inconsequential for the moments uncertainties analysis, all that is required is a detector that follows shot-noise statistics. We still provide the details for completeness sake. 


\begin{figure}
    \includegraphics[width=1\linewidth]{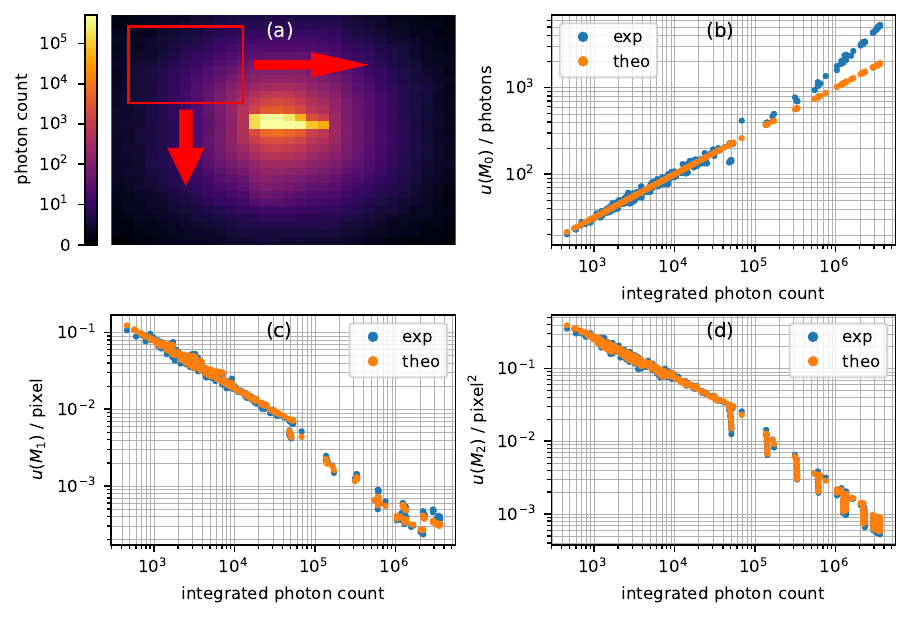} 
    \caption{Moments uncertainties measured at P06, PETRA III. (a) Frame of a single diffraction spot. The rectangle and arrows indicate the window and its variation used for the following analysis. Uncertainty of $M_0$ (b), $M_1$ (c) and $M_2$ (d) as a function of total number of photons.}
    \label{fig:p06results}
\end{figure}

The first experiment was carried out at the P06 beamline of PETRA~III in Hamburg, Germany~\cite{Schroer2010,Falkenberg2020}. The setup was implemented for x-ray diffraction with high spatial resolution similar to the one described in~\cite{Chakrabarti2022,khaliq2024}. A photon energy of 16.2~keV was selected by a double crystal monochromator. The beam was focused by Kirckpatrick Baez mirrors to 510~nm in horizontal and to 330~nm vertical direction as measured by the x-ray fluorescence signal of a Pt strip scanned through the beam. The sample, a 4H-SiC diode, was positioned on a 6-axes goniometer and the diffraction signal of the 
(0 0 0 12) reflection (see Fig.~\ref{fig:p06results}a) was collected by a Pilatus3 1M photon counting detector (DECTRIS, Switzerland) with a pixel size of 172~$\mu$m. 

For noise analysis the measurement was repeated 100 times with an exposure time of 0.1~s. We have determined the moments in terms of pixels implying $\Delta x =1$~pixel for eqs.~(\ref{eq:M0})-(\ref{eq:Mn}) and (\ref{eq:uM0})-(\ref{eq:uM2}). As mentioned above, the moments are well-defined for any experimentally obtained function. In order to determine the uncertainties of the moments as a function of both $M_0$ (i.e., total number of photons) and $M_2$ (i.e., the width of the signal), we have cut the signal to a sliding window with a size of 10 by 10 pixels, which is indicated by the red box in Fig.~\ref{fig:p06results}a. The signal in the window was vertically summed resulting in an experimentally observed one-dimensional function. The interval of total photon counts (unevenly) covered was approximately $5\cdot 10^{2}$ to $3.5\cdot 10^{6}$, while the widths of the signals (i.e., $\sqrt{M_2}$) varied from 0.38 pixels to 2.84 pixels. The experimental uncertainties were determined by the standard deviation of the moments of the individual 100 frames (i.e., eqs.~(\ref{eq:M0})-(\ref{eq:Mn})). The theoretical uncertainties were determined by the average of the experimental moments and eqs.~(\ref{eq:uM0})-(\ref{eq:uM2}).

Figure~\ref{fig:p06results}b compares the experimentally to the theoretically determined uncertainties for $M_0$ with a correlation coefficient of $r=0.984$. The deviation for high photon counts is due to dead time effects for high count rates~\cite{Trueb2015}, which was up to $3.5\cdot 10^{7}$ photons per second and per pixel. Figure~\ref{fig:p06results}c shows the experimental and theoretical uncertainties for $M_1$ with $r=0.996$. Some smaller deviations at high count rates are visible. However, it should be noted that the \first moment can be determined with an uncertainty below $\frac{1}{1000}$th of a pixel. Figure~\ref{fig:p06results}d demonstrates an equally high correlation of $r=0.995$ between the experimentally determined and theoretically predicted uncertainties of the \second moment. The small vertical sections are due to variations in the \fourth moment as the data window varies over the diffraction peak. Please note that the uncertainties for the \first and \second moments are much less affected by dead time effects than the \zeroth moment.


\begin{figure}
    \centering    
    \includegraphics[width=1\linewidth]{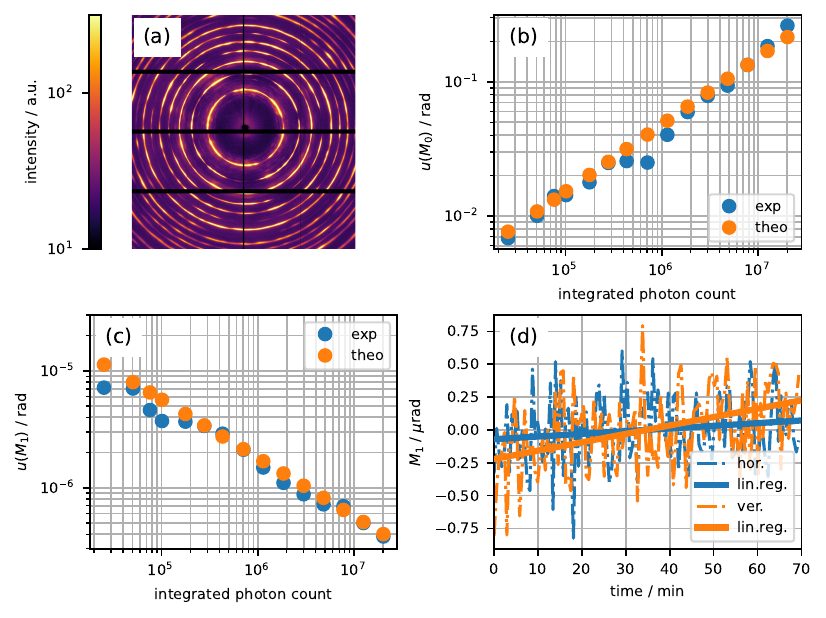} 
    \caption{Sensitivity of x-ray diffraction experiments at ID11 (ESRF). (a) Diffraction pattern of a martensitic steel sample acquired at the nanofocus station. Experimental and theoretical uncertainties of the \zeroth moment (b) and the \first moment (c) as a function of measured photon counts of the 321 diffraction ring with a Bragg angle of $\theta = 6.685$\textdegree{}. (d) Horizontal and vertical angular setup stability measured over 70~min.}
    \label{fig:id11results}
\end{figure}

The second experiment was carried out at the nanofocus station of the ID11 beamline at the ESRF in Grenoble, France~\cite{Wright2020}. The horizontally aligned Si (111) double Laue monochromator was provided a photon energy of about 70~keV. The x-ray beam was shaped by 32 AL lenses for collimation and adjusted in size by perpendicular slits to 10~$\mu$m in the horizontal and 100~$\mu$m in the vertical direction. The sample was martensitic steel with a diameter of 1~mm. The diffraction signal (see Fig.~\ref{fig:id11results}a) was collected by a photon counting Eiger2 X CdTe 4M detector (Dectris, Switzerland) about 0.3~m downstream of the sample. Calibration of the setup geometry was performed with a CeO$_2$ calibrant and pyFAI~\cite{Kieffer2020}.

Two scans have been performed. First, the diffraction pattern was measured with an exposure time of 2~ms for 15,800 times. The diffraction data was normalized and re-binned over repeated measurements to realize 20 independent instances of different effective exposure times ranging from 2~ms to 1.6~s. Each instance of the resulting diffraction patterns was then azimuthally integrated over the entire 321 ring using pyFAI to provide 1d diffraction curves. In order to retrieve the diffraction curves in terms of photon counts (and not "arbitrary" intensities) the diffraction patterns were normalized by $\sin\theta_B / \Delta \eta$ with $\theta_B$, the Bragg angle, and $\Delta \eta$, the azimuthal range (here: 360\textdegree{}) during azimuthal integration. This ensures that the \zeroth moment corresponds to the integrated photon count, i.e. $M_0 / \Delta\theta = N_{\mathrm{photons}}$. Experimental uncertainties for the \zeroth and \first moment were then calculated by the standard deviation over instances of resulting moments. Figures~\ref{fig:id11results}b and c demonstrate an excellent agreement between the experimental and theoretical uncertainties of $M_0$ and $M_1$ with correlations coefficients of $r=0.99$ and $r=0.97$, respectively.

Further, the angular stability of the setup was investigated with a second scan. This involved the repeated measurement of a diffraction pattern with 5~s exposure time and a 30~s break, which covers a time span of about 70~min. The resulting \first moments of the integrated diffraction patterns over 45\textdegree{} in horizontal and vertical direction are shown in Fig.~\ref{fig:id11results}d. Linear regression was used to provide upper bounds of the angular drift, which was 0.12 $\mu$rad/h in the horizontal and 0.38 $\mu$rad/h in the vertical direction. Please note that a sensitivity of the \first angular moment below $1~\mu$rad has been experimentally demonstrated.


\begin{figure}
    \centering
    \includegraphics[width=1\linewidth]{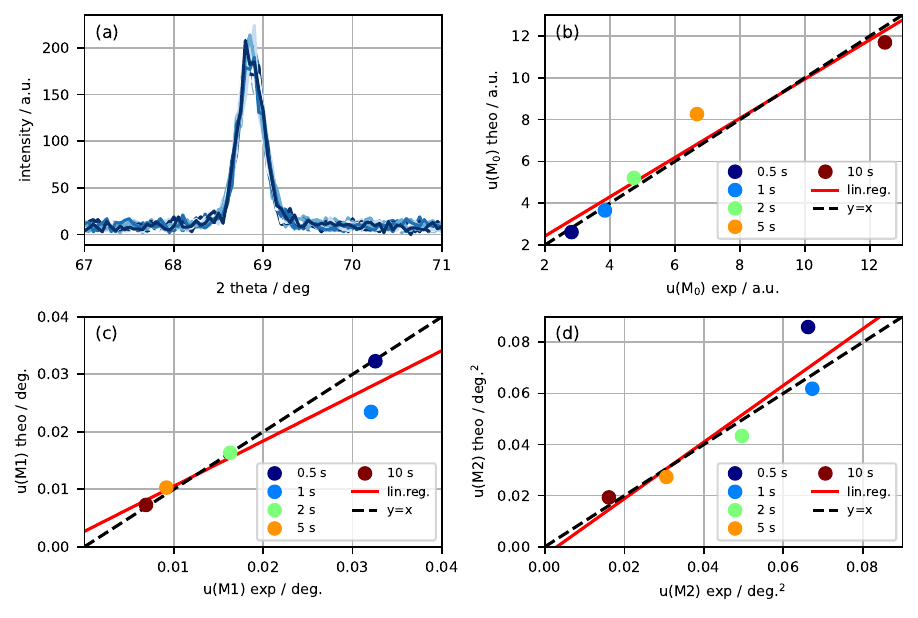}
    \caption{Uncertainties of moments measured with a laboratory x-ray diffraction setup. (a) 30 diffraction curves of the 110 reflection of a $\alpha$-Fe powder sample measured with an exposure time of 0.5~s. Scatter plots of resulting experimental and theoretical uncertainties for the \zeroth moment (b), \first moment (c) and \second moment for several exposure times. Linear regression lines are shown in red, while the $y=x$ line is shown as dashed black.}
    \label{fig:liehr}
\end{figure}

The third experiment was carried out using a laboratory x-ray diffraction setup at the University of Kassel (Germany). The four cycle  diffractometer, manufactured by HUBER, was equipped with a Chromium X-Ray tube operated at an acceleration voltage of 35~kV and an electron current of 30~mA. Furthermore, a pinhole collimator with a diameter of 1~mm and a length of 112~mm was used to define the primary beam path. At the secondary beam path a slit system with a maximum divergence slit of 0.5\textdegree{}, a K$_\beta$ filter and a standard scintillator detector was installed. The 110 diffraction peak of a standard Fe powder was collected stepwise. The scan covered an angular interval of $2\theta = \pm2$\textdegree{} with exposure times of 0.5~s, 1~s, 2~s, 5~s and 10~s to realize varying number of collected photons. For each exposure time the scan was repeated 30 times to measure the uncertainties (see Fig.\ref{fig:liehr}a). Since the utilized detector does not necessarily obey photon statistics the so-called Fano factor~\cite{Fano47} was determined for each $2\theta$ data point by calculating 
\begin{equation}
    \frac{\mathrm{std}^2(f)}{\mathrm{mean}(f)} \stackrel{\text{shot \,noise}}{=} 1,
\end{equation}
which according to eq.~(\ref{eq:photon_shot_noise}) equals 1 for a signal $f$ that is affected only by photon shot noise. Here, the average ratio over scan points was 1.06, which was considered sufficient for assuming dominating photon shot noise. We found a very good to excellent agreement between theoretically predicted and experimentally determined uncertainties for the \zeroth moment (Fig.~\ref{fig:liehr}b) with a correlation coefficient of $r=0.97$, for the \first moment (Fig.~\ref{fig:liehr}c) with a correlation coefficient of $r=0.95$ and for the \second moment (Fig.~\ref{fig:liehr}d) with a correlation coefficient of $r=0.92$.


\section{Discussion and conclusions}

The photon shot noise limited uncertainties of angular moments have been theoretical derived and compared to experimentally determined values for three different setups. Very good to excellent agreements have been found, where correlation coefficients ranged from $r=0.92$ to $r=0.996$. Sensitivities of the first moment below $1/1000$th of a detector pixel and below $1\mu$rad have been observed, which underline the sub-pixel precision of this type of data analysis.

The validated formulas for the uncertainties (i.e., eqs.~(\ref{eq:uM0})-(\ref{eq:uM2})) can now be used to rapidly determine expected sensitivities from single diffraction frames. For example, a  frame taken from the ID11 data with an exposure time of 20~ms and with the 321 reflection integrated over an azimuthal range of (-45\textdegree{},+45\textdegree{}) resulted into a diffraction curve with approximately 65,000 photons and a \second moment of $2\cdot 10^{-6}\,\mathrm{rad}^2$. The corresponding sensitivity for the \first moment according to eq.~(\ref{eq:uM1}) is approximately 6~$\mu$rad, which is in agreement with Fig.~\ref{fig:id11results}c. The provided formulas can also be used to significantly speed up numerical simulation.

\appendix
\section{Derivation of uncertainties}

In the following the details of the derivation of the uncertainties of the first three moments subjected to photon shot noise are given. For this it is convenient to first calculate the following partial derivatives, starting with the \zeroth moment defined in eq.~(\ref{eq:M0}):
\begin{equation}\label{eq:partM0}
    \partial_{f_j}\, M_0 = \partial_{f_j}\, \sum_i f_i \, \Delta x = 
    \sum_i \partial_{f_j} f_i\, \Delta x = \sum_i \delta_{ij} \Delta x = \Delta x,
\end{equation}
where $\partial_{f_j} f_i = \delta_{ij}$, Kronecker's $\delta$ was used. The corresponding partial derivative for the \first moment, given in eq.~(\ref{eq:M1}), is:
\begin{equation}
    \partial_{f_j}\, M_1 = \partial_{f_j}\, \frac{\sum_i \, x_i \, f_i \, \Delta x}{M_0} = \frac{M_0 x_j \Delta x-\Delta x \sum_i x_i f_i \Delta x}{{M_0}^2},
\end{equation}
where eq.~(\ref{eq:partM0}) was used and which can be simplified to
\begin{equation}\label{eq:partM1}
    \partial_{f_j}\, M_1 = \frac{M_0 x_j \Delta x-\Delta x M_0 M_1}{{M_0}^2} = 
    \frac{\Delta x}{M_0} \left( x_j - M_1 \right).
\end{equation}
The derivative of the  second moment, i.e. eq~(\ref{eq:Mn}) with $n=2$, requires a bit more work:
\begin{equation}
    \partial_{f_j}\, M_2 = \partial_{f_j}\, \frac{\sum_i \, \left(x_i -M_1\right)^2\, f_i \, \Delta x}{M_0}.
\end{equation}
Applying the derivative leads to
\begin{equation}
    \partial_{f_j}\, M_2 = \frac{M_0 \partial_{f_j} \left( \sum_i \left( x_i-M_1 \right)^2 f_i \Delta x\right) - \Delta x \sum_i \left( x_i - M_1\right)^2 f_i \Delta x 
    }{{M_0}^2},
\end{equation}
where again eq.~(\ref{eq:partM0}) was used. Identifying the occurring term $M_0 M_2$ in the last summand and using eq.~(\ref{eq:partM1}) yields:
\begin{equation}
    \partial_{f_j}\, M_2 = \frac{\left(x_j-M_1\right)^2\Delta x
    +  \sum_i f_i \Delta x \partial_{f_j} \left( x_i-M_1 \right)^2
    - \Delta x M_2
    }{{M_0}}.
\end{equation}
The remaining term with the derivative is:
\begin{equation}
    -2 \frac{\Delta x}{M_0} \left(x_j-M_1\right) \sum_i f_i \Delta x  \left( x_i-M_1\right),
\end{equation}
where it is straightforward to show that the occurring sum:
\begin{equation}
    \sum_i x_i f_i \Delta x - M_1 \sum_i f_i \Delta x = M_0 M_1 - M_1 M_0 = 0
\end{equation}
cancels to zero. Thus, the partial derivative of the \second moment is given by:
\begin{equation}\label{eq:partM2}
    \partial_{f_j}\, M_2 = \frac{\Delta x}{M_0}\left( \left(x_j-M_1 \right)^2 -M_2\right)
\end{equation}

Now, using the uncertainty in photon counting measurements, i.e. eq.~(\ref{eq:photon_shot_noise}), and the formula for error propagation, i.e., eq.~(\ref{eq:error_propagation}) the uncertainties of the first three moments are straight forward to calculate. Beginning with the \zeroth moment and using eq.~(\ref{eq:partM0}) retrieves the desired result:
\begin{equation}
    u(M_0)^2 = \sum_j \Delta x^2 f_j = \Delta x M_0.
\end{equation}
For the \first moment utilizing eq.~(\ref{eq:partM1}) there is:
\begin{equation}
    u(M_1)^2 = \sum_j \frac{\Delta x^2}{{M_0}^2} \left( x_j-M_1\right)^2 f_j
    = \frac{\Delta x \,M_2}{M_0}.
\end{equation}
Finally, the uncertainty of the \second moment is:
\begin{equation}
    u(M_2)^2 = \frac{\Delta x^2}{{M_0}^2} \sum_j  \left(\left( x_j-M_1\right)^2-{M_2} \right)^2 f_j,
\end{equation}
which can be rearranged to:
\begin{equation}
    u(M_2)^2 = \frac{\Delta x^2}{{M_0}^2} \sum_j \left( \left( x_j-M_1\right)^4-2\left( x_j-M_1\right)^2 {M_2}^2 + {M_2}^2  \right) f_j,
\end{equation}
where the \fourth moment, i.e. eq.~(\ref{eq:Mn}) with $n=4$, occurs. This can be further simplified to:
\begin{equation}
    u(M_2)^2 = \frac{\Delta x}{{M_0}} \left( M_4-2 M_2 {M_2} + {M_2}^2  \right) 
    = \frac{\Delta x}{{M_0}} \left( M_4-{M_2}^2 \right).
\end{equation}
Quod erat demonstrandum.



\ack{Acknowledgements}

We acknowledge DESY (Hamburg, Germany), a member of the Helmholtz Association HGF, for the provision of experimental facilities. Parts of this research were carried out at PETRA III and we would like to thank the beamline staff for assistance in using the P06 beamline. Beamtime was allocated for proposal I-20231255. Further, we acknowledge the European Synchrotron Radiation Facility (ESRF) for provision of synchrotron radiation facilities under proposal number MI-1498 and we would like to thank Jonathan Wright for assistance and support in using beamline ID-11.  Parts of these investigations were funded by the ErUM-Pro programme (grant number 05K22PS2) of the German Federal Ministry of Education and Research (BMBF). We acknowledge the Institute of Mechanical Engineering of the University of Kassel for providing samples and measurement capacity.



\referencelist{iucr}

\end{document}